\documentclass[prb,twocolumn,amsmath,amssymb]{revtex4}
\usepackage{graphicx}
 
\begin{document}

\title{Weak phase stiffness and mass divergence of superfluid in underdoped cuprates}
\author{Yucel Yildirim, and Wei Ku}
 
%
%

\address{Condensed Matter Physics and Material Science Department, Brookhaven National Laboratory, Upton,
NY 11973-5000,U.S.A.}

\maketitle

\vskip 0.5 in

{\bf
Despite more than two decades of intensive investigations, the true nature of high temperature (high-$T_c$) superconductivity observed in the cuprates remains elusive to the researchers.  In particular, in the so-called `underdoped' region, the overall behavior of superconductivity deviates $qualitatively$\cite{EK,Lee-Wen,UD1,UD2,UD3,UD4,UD5,UD6,UD7,UD8,UD9,UD10} from the standard theoretical description pioneered by Bardeen, Cooper and Schrieffer (BCS).\cite{BCS}  Recently, the importance of phase fluctuation of the superconducting order parameter\cite{EK,Uemura,orderpar1,orderpar2,ChargedLatticeBoson} has gained significant support from various experiments.\cite{Samuele,Broun,Khasanov,Hardy}  However, the microscopic mechanism responsible for the surprisingly soft phase remains one of the most important unsolved puzzles.  Here, opposite to the standard BCS starting point, we propose a simple, solvable low-energy model in the strong coupling limit, which maps the superconductivity literally into a well-understood physics of superfluid in a special dilute bosonic system of local pairs of doped holes.  In the prototypical material (La$_{1-\delta}$Sr$_\delta$)$_2$CuO$_4$, without use of any free parameter, a $d$-wave superconductivity is obtained for doping above $\sim 5.2\%$, below which unexpected incoherent $p$-wave pairs dominate.  Throughout the whole underdoped region, very soft phases are found to originate from enormous mass enhancement of the pairs.  Furthermore, a striking mass divergence is predicted that dictates the occurrence of the observed quantum critical point.  Our model produces properties of the superfluid in good agreement with the experiments, and provides new insights into several current puzzles.  Owing to its simplicity, this model offers a paradigm of great value in answering the long-standing challenges in underdoped cuprates.
}

{\bf Introduction}

Considering the enormous amount of research activities devoted to the problem of high-$T_c$ superconductivity, it is hardly an exaggeration to regard it as one of today's most important unsolved problems in physics.  In particular, in the underdoped region where only small number of `hole' carriers exist in the system, both the low-temperature superconducting phase and the `normal state' above the transition temperature $T_c$ demonstrate exotic properties\cite{EK,Lee-Wen,UD1,UD2,UD3,UD4,UD5,UD6,UD7,UD8,UD9,UD10,toni} inconceivable from the standard BCS theory.  Even if one focuses only on the properties of the superfluid, as in this study, still lacking is a comprehensive and coherent description capable to simultaneously account for, for example, its $d$-wave symmetry, its rapid thermal depletion, its very soft phase, and its dramatic destruction near the 5.2\% doping.

Perhaps the most exotic property of underdoped cuprates is the observation of strong diamagnetism\cite{dia1,dia2,dia3}, one of the signatures of superconductivity, at temperature much higher than $T_c$.  This important observation suggests the existence of ``local'' super-current capable of screening the magnetic field in the non-superconducting state, and thus has been taken as direct evidence of `preformed pairs'\cite{UD6,pairs1,pairs2}.  Following the pioneering observation of proportionality between $T_c$ and zero-temperature superfluid density\cite{Uemura}, $n_s(T=0)$, Emery and Kivelson\cite{EK} made a general macroscopic argument:  At low carrier density£¬ the phase of the complex superconducting order parameter is soft enough such that its fluctuation can destroy superconductivity before its magnitude vanishes.  The observed $T_c$ is thus no longer controlled by the pairing strength $U$ as in any BCS-like mean-field theories, but instead by the phase `stiffness'.

This important realization, however, left behind the most essential microscopic question: ``Why is the phase in cuprates so soft?''  As demonstrated in Fig.~1a for the prototypical high-$T_c$ cuprates (La$_{1-\delta}$Sr$_\delta$)$_2$CuO$_4$ (LSCO), the phase stiffness $V_0=\hbar^2n_s(0)a/4m^*$, evaluated using realistic $c$-axis lattice constant $a=6.6\AA$, effective mass $m^*$ from twice the experimental mass of the holes, $m_h^*\sim 4.8m_e$,\cite{BareMass,EffectiveMass} and $n_s(0)=\delta/\Omega_0$, doping level $\delta$ per unit cell volume $\Omega_0$, leads to a $T_{\theta}\approx 2.2V_0$\cite{EK} much higher than the experimental observation.  Obviously, a microscopic explanation to the very soft phase in the underdoped region is one essential piece of the puzzle.

A related outstanding issue is the nature of the quantum critical point (QCP) at $\delta=\delta_{QCP}\sim 5.2\%$, the point where superconducting phase vanishes at zero temperature.  As opposed to the above estimation of $T_{\theta}\propto\delta$, the actual $T_c$ diminishes rapidly to zero at $\delta_{QCP}$ without an immediate appearance of a competing phase.  A simple picture is still unavailable on exactly what happens to the superfluid near the QCP that suppresses $T_c$ so dramatically, and why the superconductivity ceases to exist below this point even at zero temperature.


Here, we propose a simple low-energy model to address these crucial questions in the underdoped region, by starting from the strong coupling limit opposite to the BCS approach.  The superconductivity is then mapped literally into a well-understood problem of superfluid of a special dilute bosonic gas of local pairs of doped holes.  The model is found to lead to the above puzzling properties of the prototypical material (La$_{1-\delta}$Sr$_\delta$)$_2$CuO$_4$ without use of any free parameter.  Specifically, a strong competition with incoherent $p$-wave symmetry is found to enormously enhance the mass of the $d$-wave pairs, explaining the key puzzle of the very soft phase of the complex order parameter.  Furthermore, a striking mass divergence is predicted that dictates the occurrence of quantum critical point observed at $\sim 5.2\%$ doping.  Our model describes most properties of the superfluid very well compared with the experiments, and provides new insights into several current puzzles.  This simple and solvable model offers a useful paradigm answering the long-standing challenge of describing the underdoped cuprates at low temperature.

{\bf Model in the strong coupling limit}

Our simple model is derived in the following steps.  First, the effective one-particle kinetics of the low-energy doped holes is extracted from the experimental spectral functions of the `normal state' at temperature slightly higher than $T_c$, the so-called `pseudo-gap' phase.  Next, we consider the implications of strong coupling limit and define a bosonic description of the paired holes.  Finally, the pivoting motion of the pairs is derived presenting an extended hard-core constraint.

Figure 1b shows the dispersion of the main features in the measured spectral functions by angular-resolved photoemission spectroscopy (ARPES)\cite{ExpArpes1,ExpArpes2,ExpArpes3,ExpArpes4}, and our theoretical fit via a simple tight-binding Hamiltonian.  One notices immediately the well-known observation that the dispersion is not rigid against the doping level.  The close resemblance to the previously published $t$-$J$ model solutions\cite{Weiguo} in Fig.~1c suggests that this strong doping-dependent renormalization originates primarily from the competition between the bare kinetic energy and the anti-ferromagnetic (AF) interaction.  As a consequence, the resulting fully-dressed first, second and third neighboring hopping parameters, $t$, $t^\prime$ and $t^{\prime\prime}$, of the holes (Fig.~1d) demonstrate strong doping dependence as well.  Interestingly, as $\delta$ decreases, $t^{\prime\prime}$ is found to increase steadily approaching the value of $t^\prime$, and then rapidly exceeds $t^\prime$ {\it right at $\delta_{QCP}$!}  This is apparently not a coincidence, and reveals an important clue to the nature of QCP to be discussed below.  As a reference, Fig.~1e also shows the effective mass of the doped holes, $m^*_h$, in three directions.  In agreement with the experimental observation\cite{EffectiveMass}, $m^*_h$ (red line) is about constant for $\delta>5.2\%$.

Note that the resulting hopping parameters should be properly understood as only convenient representations of the average effective kinetics of one-particle propagator under the full renormalization of the many-body interactions.  They certainly do not contain explicit information of two-particle interaction nor any decoherent processes that broaden the spectrum.  Since the fully dressed $t$ is negligibly small, understandable from the strong AF correlation, it will be dropped from our further analysis.

Now, taking the observation of strong diamagnetism at $T>T_c$ as a direct evidence of existence of locally coherent pairs, let's consider the strong coupling limit, where the pairing strength between doped holes is stronger than the renormalized kinetic energy, such that at low temperature $T\leq T_c$ the doped holes are mostly paired with other doped holes.  This non-conventional condition is well realized in the underdoped region, as made explicit from the well-established Gutzwiller projection\cite{Gutzwiller} with strong AF correlation, and/or with the formation of bi-polaron\cite{biPolaron}.  Furthermore, since it is unlikely that doped holes can doubly occupy the same site in a weakly doped AF Mott insulator, the doped holes are then expected to be mostly nearest neighbor to each other under such a strong coupling.  It is thus convenient to employ a bosonic picture consisting of pairs of tightly-bound first-neighboring doped holes, $b_{ij}^\dagger=c_{i \uparrow}^\dagger c_{j \downarrow}^\dagger$ located at site $i$ and $j$ with opposite spin.  In essence, the low-energy physics of the underdoped system is then mapped to a dilute bosonic gas, a well understood problem that can be solved quite accurately.

The strong coupling limit is the exact opposite to the standard BCS theory whose validity relies on the weakness of pairing interaction in comparison to the kinetic energy.  Not surprisingly, the key low-energy physics is quite different in this case.  For example, consistent with Anderson's ``mammoth vs. mouse'' argument\cite{AndersonGlue}, typical concerns about the pairing mechanism (i.e. exchange of glue particles like the phonon or magnon) is irrelevant at $T\leq T_c$, since pairing is so much stronger than $T_c$ that the high-energy depairing process is of no direct relevance to the properties of the superfluid itself.  Instead, the key factor is the degree of phase coherence, the focus of this study.  Interestingly, the removal of interaction strength from the low-energy physics leaves the effective kinetic energy the only relevant energy scale, allowing a simple and generic description shown below.

As another example, the superconducting pairs are typically considered as pairs of electrons (or holes) made of quasi-particles with opposite momentum, but here a tightly bound pair consists of superposition of the $whole$ momentum space.  In the strong coupling limit, holes being part of a pair cannot be considered as quasi-particles at all, due to the strong pairing interaction.  As shown elsewhere\cite{paper2}, the very low-energy quasi-particles observed by ARPES\cite{UD8,arpes2,arpes3,arpes4} or STM\cite{stm1} are in fact not the typical Bogoliubov quasi-particles that get knocked out of the pairs.  Similarly, the typical decay of superconducting pairs via depairing into two low-energy quasi-particles\cite{Lee-Wen} is negligible at low temperature here\cite{paper2}.  All these seemingly anti-intuitive physics only reflects the well-known fact that a strongly interacting system may not be simply considered as a collection of weakly interacting quasi-particles.  {\it Extreme caution} should thus be exercised in such limit before applying existing knowledge and intuition built on the BCS theory.

Most importantly, the strong binding between the paired holes modifies significantly the kinetics of the boson, as demonstrated in Fig.~2.  Consider a single pair (blue and red filled diamonds) located in the fermion lattice in Fig.~2a.  The hopping of each hole is now constrained by the requirement of staying paired.  As a result, only three potential destinations (empty diamonds) for each hole are allowed, two via second neighbor hopping, $t^\prime$, one via third neighbor hopping, $t^{\prime\prime}$.  (Higher order processes involving temporary pair breaking are negligible in the strong coupling limit.)  Converted to the bosonic description in Fig.~2b, one finds two inequivalent bosonic sites in a square lattice, each can hop to four first neighboring sites with $t^\prime$, but only \textit{two} second neighboring sites with $t^{\prime\prime}$ along their designated direction (denoted by the direction of the ellipsoids).  This incomplete lattice forms the building block or our model.  The resulting bosonic Hamiltonian reads:
\begin{equation}
  H = \sum_{ll^\prime} t_{ll^\prime}b_l^\dagger b_{l^\prime} + h.c. [+ constraint]
\end{equation}
where $l$ denotes the bosonic lattice sites and $t_{ll^\prime}$ has value $t^\prime$ for the first neighbors and $t^{\prime\prime}$ for the designated second neighbors as illustrated in Fig.~2b.

It is essential to note that the bosonic operators, $b$'s are under a serious `extended hardcore constraint': $b_{ij}^\dagger b_{i'j'}^\dagger=0$ if $i=i'$ or $j=j'$, inherited from the Pauli exclusion principle of the original fermion operators $c_{i\sigma}^\dagger c_{i\sigma}^\dagger=0$.  In the bosonic lattice shown in Fig.~2b, this constraint forbids occupation of the six potential hopping destinations of each boson by another boson.  In other words, the real-space pairs are correlated via a ``polite'' policy that no one is to get into other's immediate hopping path.  This constraint can be considered as an infinite short-range repulsion that dominates the inter-bosonic interactions in most cases, and is responsible for stabilizing our bosonic system against potential phase separation\cite{NoPhaseSep}.  Also note that our model can actually be derived equivalently through the use of full one-particle propagator diagrammatically, but the derivation presented here is more intuitive and reveals more insights in the following analysis.

{\bf Soft phase and mass divergence of $d$-wave pairs}

Following the standard approach to dilute bosonic systems, let's first examine the solution of the kinetic part of the Hamiltonian, as it turns out to be able to answer all the main questions of this study already.  For convenience, a unit cell is chosen to include four boson sites as shown in Fig.~2b.  This choice explicitly allows one s-, two p-, and one $d$-wave superposition within the unit cell, and equates the doping level per unit cell of boson to that of the standard fermion lattice.  Fig.~3a illustrates the resulting bosonic band structure (band energy $\epsilon_k$ vs. crystal moment $k$) with doping level ($\delta=15\%$) above $\delta_{QCP}$.  Interestingly, the lowest energy band is composed of $d$-wave symmetry (red color) with a single minimum, where Bose-Einstein condensate (BEC) would take place at low temperature.  One thus expects the superfluid behavior (boson flowing without dissipation) once the constraint (interaction) is enabled.  That is, the $d$-wave superconductivity would appear in the form of bosonic superfluidity in the presence of a true BEC.

The real-space point of view provides further insights into the low-energy Hilbert space of the solution.  As shown in Fig.~2a, the local Wannier function, calculated as Fourier transform of the Bloch functions of the lowest band, has clear $d$-wave symmetry with nodes along the $(\pi,\pi)$ directions of the standard Fermion lattice, in agreement with the experimental observations.\cite{UD8,arpes2,arpes3}  (Recall that our bosonic lattice is 45-degree rotated from the fermion lattice.)  That is, the $d$-wave symmetry is entirely a local phenomenon unrelated to the phase coherence.  Therefore, the unconventional $d$-wave symmetry only serves as a ``form factor'' of experiments that observe the pairs via their coupling to the fermions.  It is, however, {\it completely irrelevant} to the general behavior of superfluid itself.  On the other hand, as far as phase coherence's concern (how bosons with the ``shape'' of the Wannier function aligning their phases across the system), the zero momentum of our BEC is of no difference from that of the standard BCS $s$-wave superconductivity.

Our resulting local $d$-wave symmetry is {\it entirely driven by the screened kinetic energy}, consistent with a previous study of the $t$-$J$ model.\cite{Elbio2}  Indeed, the positive sign of $t^\prime$ of the local pair prefers energetically opposite sign of the wave function across first neighbors, favoring a $d$-wave symmetry.  On the other hand, the positive sign of $t^{\prime\prime}$ favors opposite sign across the second neighbors, thus $p$-wave symmetry (denoted via green color).  This produces dominant {\it incoherent local p-wave pairs at $\delta <\delta_{QCP}$}, as shown in Fig.~3c.

The competition between $d$-wave and $p$-wave leads to a crucial physical effect, namely a significant mass enhancement.  Even near the optimal doping ($\delta\approx15\%$), the comparable value of $t^{\prime\prime}$ and $t^\prime$ leads to a large effective mass $m^*=(\hbar^2/l^2) d^2\epsilon_k/dk^2 \approx 12m^*_h \approx 59m_e$ (with lattice constant $l$), corresponding to a penetration depth $\lambda=\sqrt{{{m^*c^2}\over {4\pi e^2 n_s}}} \approx 7000 \AA$, in reasonable agreement with the experimental value \cite{LambdaExp} (taking $n_s\sim \delta / 2$ per Cu atom).  Furthermore, as $\delta$ approaches $\delta_{QCP}$, $t^{\prime\prime}$ grows to the value of $t^\prime$, reducing the separation of the $d$-band and the $p$-band, and in turn flattening the $d$-band.  The effective mass of the $d$ band thus increases significantly (Fig.~3d), and eventually diverges at the QCP, where $p$- and $d$-wave become degenerate at $t^{\prime\prime}=t^\prime$ (Fig.~3c).  Since the superfluid stiffness correlates with the inverse of the effective mass, {\it the large mass enhancement would dictate a very soft phase}.

Our results also lead to an unexpected exotic conclusion: {\it The QCP at the end of the underdoped superconductivity region is dictated by a mass divergence of the local pairs}.  With an infinite mass, the phase of local pairs becomes entirely free even within short distance, since they can no longer communicate with each other.  Even at zero temperature, this complete lost of phase coherence would suppress entirely the BEC.  Equivalently from the bosonic perspective, the infinite number of minimum in the flat band dispersion in Fig.~3b can no longer support a BEC.  Consequently, the diamagnetic response would also be strongly suppressed near the QCP due to lack of short-range coherence, in perfect agreement with the experimental observation\cite{OngNernstEffect&Diamagnetism}.

Interestingly, at $\delta<\delta_{QCP}$, the mass of the $p$-wave pair still diverges along the anti-nodal $(\pi,\pi)$ directions, dictated by the symmetry of our lattice in Fig.~2b, rendering BEC impossible as well.  It would be curious to investigate the scattering against entirely incoherent $p$-wave, instead of the impurities, as a microscopic origin of the well-known glassy behaviors found in this region.

Note that in great contrast to the mass divergence associated with the metal-insulator transition\cite{MassDivMott}, the above divergence results from the competition between kinetic effects of $t^\prime$ and $t^{\prime\prime}$.  This competition can be considered as a destructive interference resulting from the wave nature of the quantum particles.  As illustrated in Fig.~4, in comparison to the weakly bound superconducting pair in the BCS theory, a tightly bound local pair is subject to a strong interference effect, and thus reduced net kinetics (or enhanced mass).  The surprising mass divergence at the QCP can thus be viewed as a complete destructive interference that renders the hole pairs immobile.

Our predicted mass divergence near QCP is strongly supported by indirect experimental evidences.  The inverse square of the experimental penetration depth\cite{Sonier}, $\lambda^{-2}$, of YBa$_2$Cu$_3$O$_y$ (YBCO) shows non-linear behavior against doping (Fig.~5a).  More significantly, near the QCP, measurement performed on a single YBCO sample under repeated annealing to modulate the doping level\cite{Broun} cleanly shows a strongly non-linear correlation between $\lambda^{-2}$ and $T_c$ (Fig.~5b).  The zero slope at $\lambda\rightarrow 0$ in both cases can be considered indirect evidence of the mass divergence.  Nonetheless, a direct measurement of the superfluid inertia would be highly valuable to unambiguously verify our conclusion.






{\bf In-plane superfluid properties}

So far, our model has managed to address the key issues of this study without use of any free parameter.  Below we verify the applicability of our model in describing the in-plane behavior of the superfluid in comparison with the available experimental data.  To this end, we adapt numerically the well-established self-consistent $T$-matrix approach\cite{Tmatrix1,Tmatrix2} to the screened interaction and apply the Bogoliubov transformation\cite{Bogoliubov} to study the superfluid properties (see supplementary materials for detail).  Taking the extended hardcore constraint as the leading ``interaction'' corresponding to a hard-sphere radius comparable to the lattice constant $l$, the scattering length of the low-energy particle (the Wannier function) is estimated to be $a\approx l/4$.

Naturally, a true BEC also requires tunneling of the bosons out of the plane.  This is, however, not easily treated accurately due to the highly ``effectively disordered'' nature of the strongly coupled pairs in our picture.  Given the weak interlayer hopping of the holes ($t_z\sim 12meV$\cite{fermionTz}), the out-of-plane tunneling of a pair may encounter one of the following three possibilities.  First, one of the holes in the pair hops to the next plane and happens to encounter another unpaired hole to form a new pair.  This first order process, however, is unlikely due to the rare supply of unpaired holes in the strong coupling limit.  Second, one of the holes hops to the next plane, followed by its partner.  Taking the magnetic coupling $J/4\sim 33meV$\cite{ValueOfJ} as the representitive pairing interaction $U$, this second-order process gives tunneling coefficient $t_\perp=2t_z^2/U\sim 9meV$ (or smaller if other real-space attractions also contribute).  Of course this second possibility is only allowed when the spins across the plane happens to align ferromagnetically in the vicinity of the pair.  This brings the third possibility that the tunneling would be forbidden by the AF arrangement across the plane.  Obviously, such a violent (and strongly $\delta$-, $T$- and site-dependent) on/off modulation of the tunneling is expected to limit significantly the superfluid density in the out-of-plane direction, through shorter bosonic lifetime or possibly development of localization\cite{AndersonLocalization}.  This analysis also indicates that while the in-plane AF correlation might have been one of the leading factors to provide a strong pairing of holes, a strong AF correlation across the plan is actually very damaging to the formation of BEC and to the out-of-plane superfluid density.

We thus limit the following discussion of the superfluid to the in-plane properties only.  For simplicity, a fixed bosonic tunneling $t_\perp\sim 3meV$ is chosen as a parameter that resembles a rough average over the above three possibilities.  This should be understood as a crude estimate to reproduce the average degree of coherence only in the study of in-plane superfluid density.  (Such a simple average is certainly inadequate to give anything sensible for the out-of-plane superfluid density.)  All the results reported below corresponds to this single set of parameters $t_\perp$ and $a$.  At low temperature, the scattering with the BEC results in quasi-particle energy $E_k=\sqrt{\epsilon_k^2+2\gamma \epsilon}$, consisting of linear dispersion (sound propagation) at the long wave-length for energy within the characteristic scale $\gamma$.  Therefore, the flow of the pairs suffers no dissipation below the superfluid velocity, $v_s=\hbar\sqrt{\gamma/m^*}$.

Figure 5c shows the resulting phase diagram, in which $T_\theta$ is determined by solving $\int{ DOS(w)*f_B(w)dw}=n$ numerically, where $n=\delta/\Omega_0$ is the hole carrier density and $DOS(w)$ is the ``bare'' density of state corresponding to the band structure in Fig.~3.  Owing to the above choice of parameters, the resulting $T_\theta$ reproduces the experimental phase boundary quite well at the underdoped region.  (The experimental $T_c$ is expected to deviate from our $T_\theta$ near the optimal doping, where amplitude fluctuation becomes relevant.)

Moreover, all our key conclusions reflects explicitly in the superfluid density, $n_s=n-n_n$, where the normal component of the number density $n_n$ is obtained via 
$n_n={\displaystyle\lim_{v\to 0}} {<\vec{P}>_{v}.\vec{v}\over {v^{2} m^* \Omega}} = \int{ DOE'(w)f_{B}(w) dw}$ where $DOE(w)\cong \int{{d^3k \over {2\pi^3}}{\hbar^2 k_{i}^2 \over m^*}\delta(w-E_k)} = (2w\epsilon / 3(\epsilon+\gamma)) DOS(\epsilon)$ with $\epsilon=\sqrt{\gamma^2+w^2}-\gamma$.  Indeed, the zero-temperature superfluid density in Fig.~5d vanishes rapidly near the QCP, and remains zero below QCP, despite the finite density of local pairs (red dotted line.)  Interestingly, our resulting penetration depth compares quite well with the experimental trends against $\delta$ and $T_c$ (Fig.~5a and 5b.)

Several important issues concerning the temperature dependence of superfluid density can now be addressed by our results.  As shown in Fig.~5e, experimental $n_s(T)$ at $\delta\approx 7\%$ shows a approximately linear reduction at low temperature, in great contrast to the exponentially small reduction in standard $s$-wave BCS theory.  This experimental observation was originally viewed as an evidence of the $d$-wave structure of the superconducting order,\cite{Hardy} resulting from superfluid decaying into quasi-particles near the nodes, where the superconducting gap decreases to zero.  Also shown in Fig.~5e, such rapid reduction of $n_s(T)$ can be reproduced very well by our calculation.  As pointed out earlier, in the strong coupling limit assumed in this work, decay into quasi-particles is inefficient.  Instead, in spirit similar to the alternative explanation by Emery and Kivelson,\cite{EK} the rapid reduction of $n_s(w)$ in our results originates from the thermal depletion of the BEC, completely unrelated to the local $d$-wave structure.

A more serious question was raised in the recent measurement of $\lambda$ near the QCP, also shown in Fig.~5e, namely the absence of presumable quasi-2D nature of the superfluid.  The low-energy hole carriers in cuprates are well accepted to be quasi-2D.\cite{UD8,quasi2D}  Naturally, the sharp drop of $n_s(T)$ at $T\sim 0.8T_c$ for $\delta\approx 7\%$ was commonly interpreted as the signature of the so-called BKT transition\cite{BKT1,BKT2,BKT3} associated with the quasi-2D nature.  However, such rapid reduction is entirely absent near QCP.  In addition, in both 7\% and near QCP, $n_s(T)$ appears to be linear near $T_c$,
casting doubt on the quasi-2D assumption of the superfluid.

Our results again capture these features very well and thus provide a potential resolution to this important issue.  It turns out that the sharp drop of $n_s(T)$ at higher doping (e.g. 15\%) only reflects approximately the energy scale of $2\gamma(T)$, the energy scale of linear dispersion of the bosonic quasi-particles.  Throughout the underdoped region, {\it the enormous in-plane mass enhancement suppresses significantly the quasi-2D nature of the superfluid}.  Especially close to the QCP, the near divergence of the in-plane mass obviously will not support any quasi-2D behavior, even though there still exists qualitative difference in the out-of-plane direction due to its strongly disordered characteristics.


Finally, our model gains further verification from comparing experimental critical current with the intrinsic theoretical limit $j_c <= en_sv_s$.  Figure 5f shows three measurements\cite{jc1,jc2,jc3} all of which consists of an reflection point around $T\approx0.8T_c$, also produced nicely in our calculation.


As final remarks, our main findings, namely 1) kinetic origin of the $d$-wave symmetry, 2) soft phase due to mass enhancement and 3) QCP dictated by mass divergence, all depend only on the ARPES measurement and require no free parameter.  Our current study does not include the special case of the `stripe' phase\cite{TranquadaStripe}, although it should be possible to describe it as a long-range order of the local pairs, with a microscopic derivation of the anti-phase `$d$-density wave' phenomenon.\cite{dDW1,dDW2}  Our bosonic picture is built on bond-centered boson and is thus ideal for the description of the observed broken $C_4$ symmetry in STM\cite{STM_C2}.  Finally, extension of current work to include pairing beyond first neighbors (beyond the strong coupling 'limit') can be formulated in a similar manner, but such extension will not change the symmetry properties of the solution, and thus should preserve qualitatively the key findings of this work.


{\bf Conclusion}

In summary, utilizing the strong AF correlation, we propose a very simple microscopic model of strong coupling limit of superconductivity in the underdoped cuprates, where physics is evidently dominated by phase fluctuation.  The $d$-wave symmetry is found to be entirely a local phenomenon driven by the kinetic energy, and the observed superconductivity can be understood as a superfluid of a dilute bosonic gas of local pairs.  Furthermore, a severe competition between first and second neighbor hoppings leads to a colossal mass enhancement at underdoped region, which in turn produces the observed very soft phase.  The presence of quantum critical point at the end of the underdoped superconducting region is predicted to be dictated by a mass divergence of the pairs with which the phase coherence is totally lost.  Our results answer several key questions on the underdoped cuprates and appear quite adequate for describing the in-plane superconductivity in the underdoped region in a simple and coherent manner.

\acknowledgments

The authors aknowlege fruitful dicussions with Maxim Khodas and Chris Homes, and useful comments from Alexei Tsvelik and Weiguo Yin.  This work was supported by the U.S. Department of Energy, Office of Basic Energy Science, under Contract No. DE-AC02-98CH10886.

Correspondence and requests for materials should be addressed to Wei Ku (email: weiku@mailaps.org)

\newpage

\begin{figure}
\caption{Fig.~1  (a) Phase diagram of LSCO in temperature $T$ vs. hole doping ($\delta_{h}$).  $T_{MF}$ gives the $T_c$ from mean-field theory \cite{RVB}.  $T_{\theta}$ denotes the temperature of appearance of phase coherence, and $T_c$ the experimental superconducting transition temperature.  (b) Illustration of our fitting to the experimental band dispersion.  (c) Band structure of $t$-$J$ model\cite{Weiguo}.  (d) Parameters $t$, $t^\prime$ and $t^{\prime\prime}$ of holes obtained from (b).  (e) Doping dependence of carrier mass, $m^*_h$, in different directions indicated by the arrows in the inset.
}
\label{fig1}
\end{figure}

\begin{figure}
\caption{Fig.~2  (a) Illustration of kinetic hopping of a pair of holes (filled squres).  Open squres indicate the allowed destinations under the pairing constraint.  Ellipsoids denote the same processes from the perspective of the pair as a whole. (b) Illustration of bosonic kinetic processes of the paired holes.  The solid/dashed lines denote first-/second-neighbor hopping paths of strength $t^\prime$/$t^{\prime\prime}$.  The yellow area shows the region of the `extended hardcore contraint' where occupation of aother pair is forbidden.  The dotted square gives the unit cell for our calculations.
}
\label{fig2}
\end{figure}

\begin{figure}
\caption{Fig.~3  The band dispersion of kinetic-only solution at $\delta>\delta_{QCP}$(a), $\delta=\delta_{QCP}$(b), and $\delta<\delta_{QCP}$(c).  Red, green and blue color represents the local $d$-, $p$-, and $s$-wave symmetry.  Insets in (a) and (c) illustrates the symmetry of the Wannier function corresponding to the lowest set of bands, and the kinetic effects that drive the symmetry.  (d) Effective mass of the pairs, $m^*$, and the measured holes, $m^*_h$.
}
\label{fig3}
\end{figure}

\begin{figure}
\caption{Fig.~4  Illustration of heavier mass resulting from stronger binding.}
\label{fig4}
\end{figure}

\begin{figure}
\caption{Fig.~5  (a) Non-linear doping dependence of penetration depth.  (b) Non-linear temperature dependence of penetration depth.  (c) Phase diagram with our model (blue dots; other symbols identical to Fig.1a), with inset showing detail near the QCP.  (d) Doping dependence of the superfluid density.  (e) Temperature dependence of superfluid density for various doping levels.  (f) Temperature dependence of critical current.
}
\label{fig5}
\end{figure}

\newpage
\begin{figure}[thbp]
\begin{center}
\includegraphics[width=15cm,clip,angle=0]{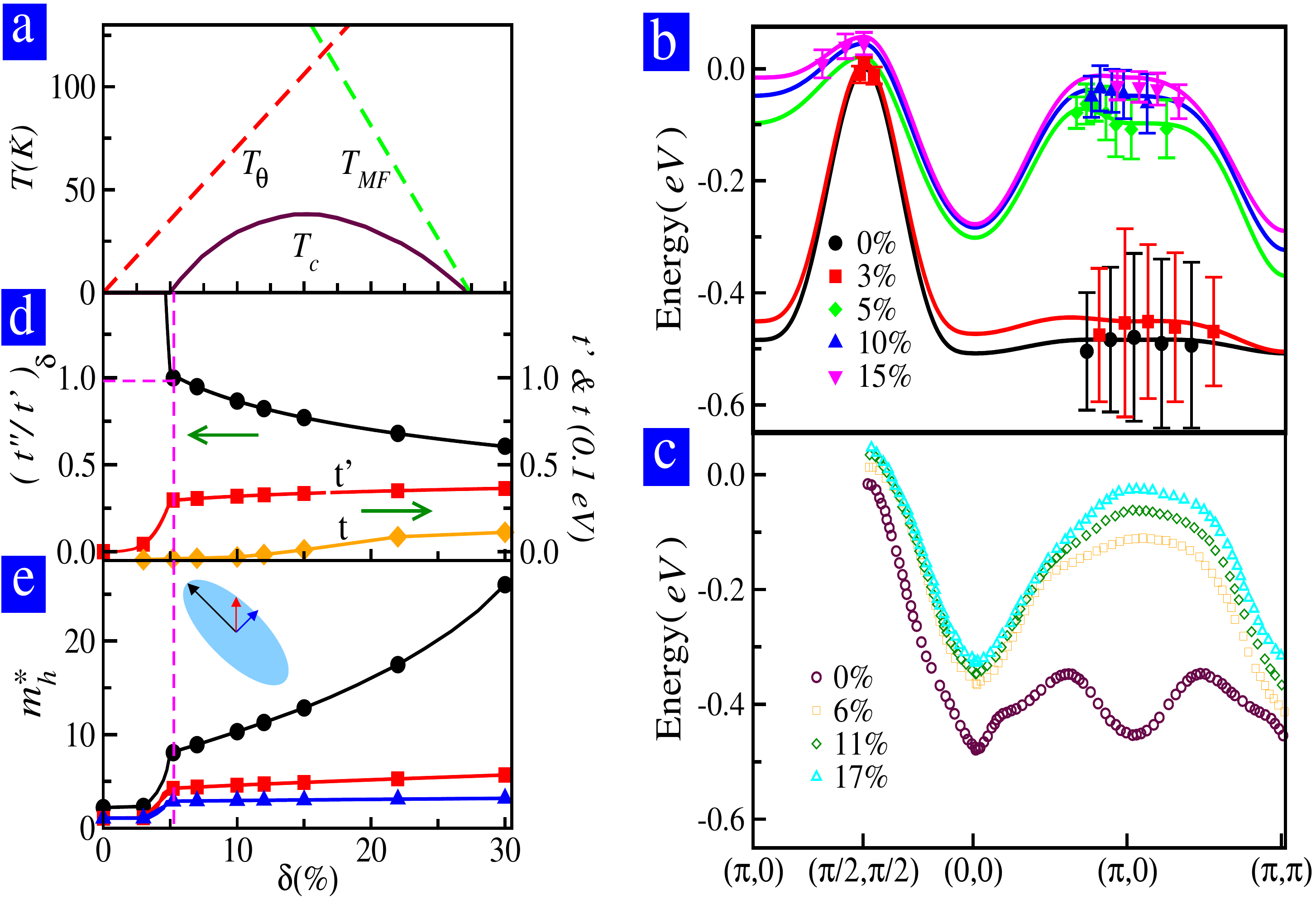}
\vskip 0.3cm
\label{F1}
\end{center}
\end{figure}

\newpage
\begin{figure}[thbp]
\begin{center}
\includegraphics[width=18.5cm,clip,angle=0]{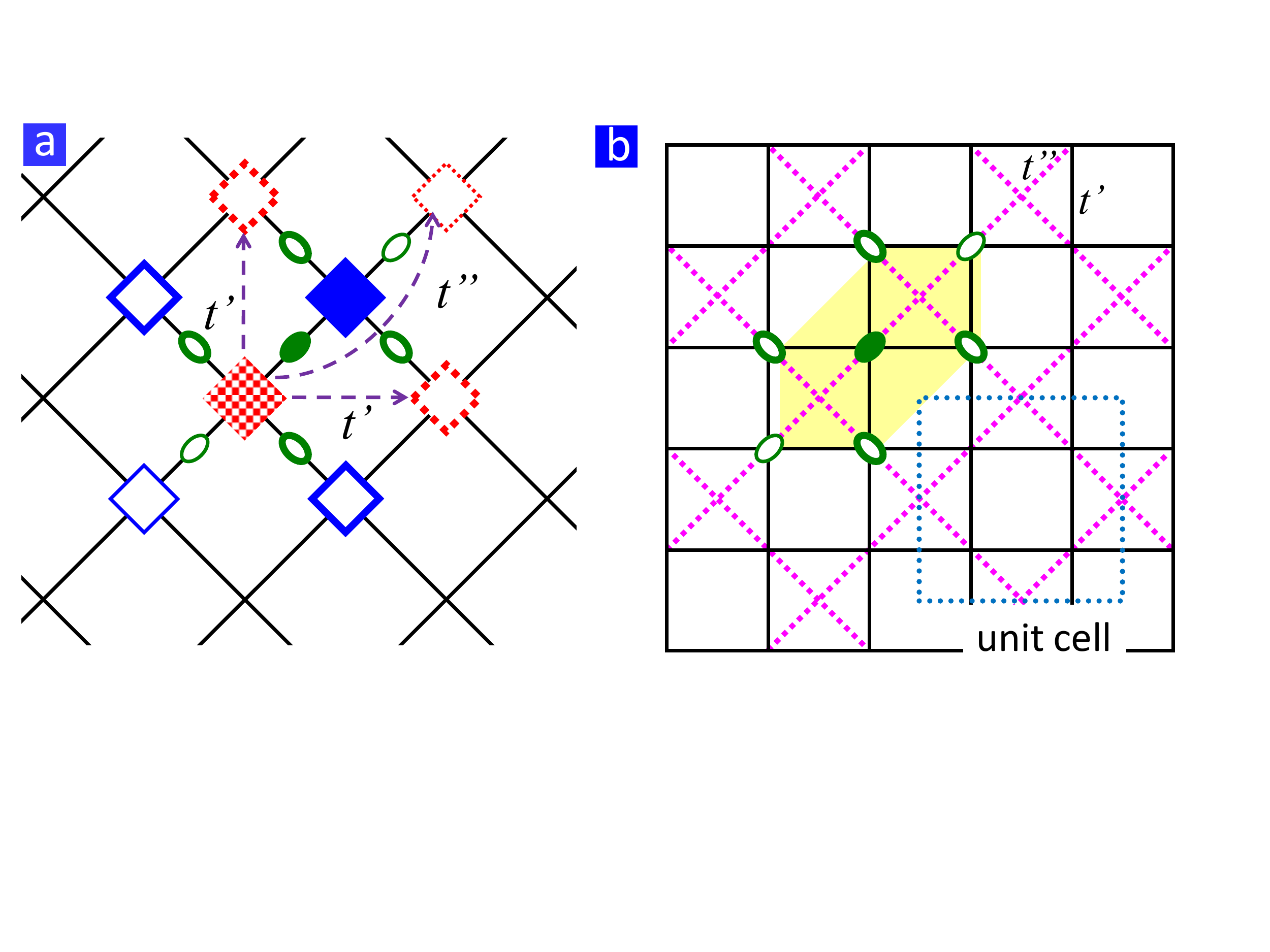}
\vskip 0.3cm
\label{F1}
\end{center}
\end{figure}

\newpage
\begin{figure}[thbp]
\begin{center}
\includegraphics[width=18.5cm,clip,angle=0]{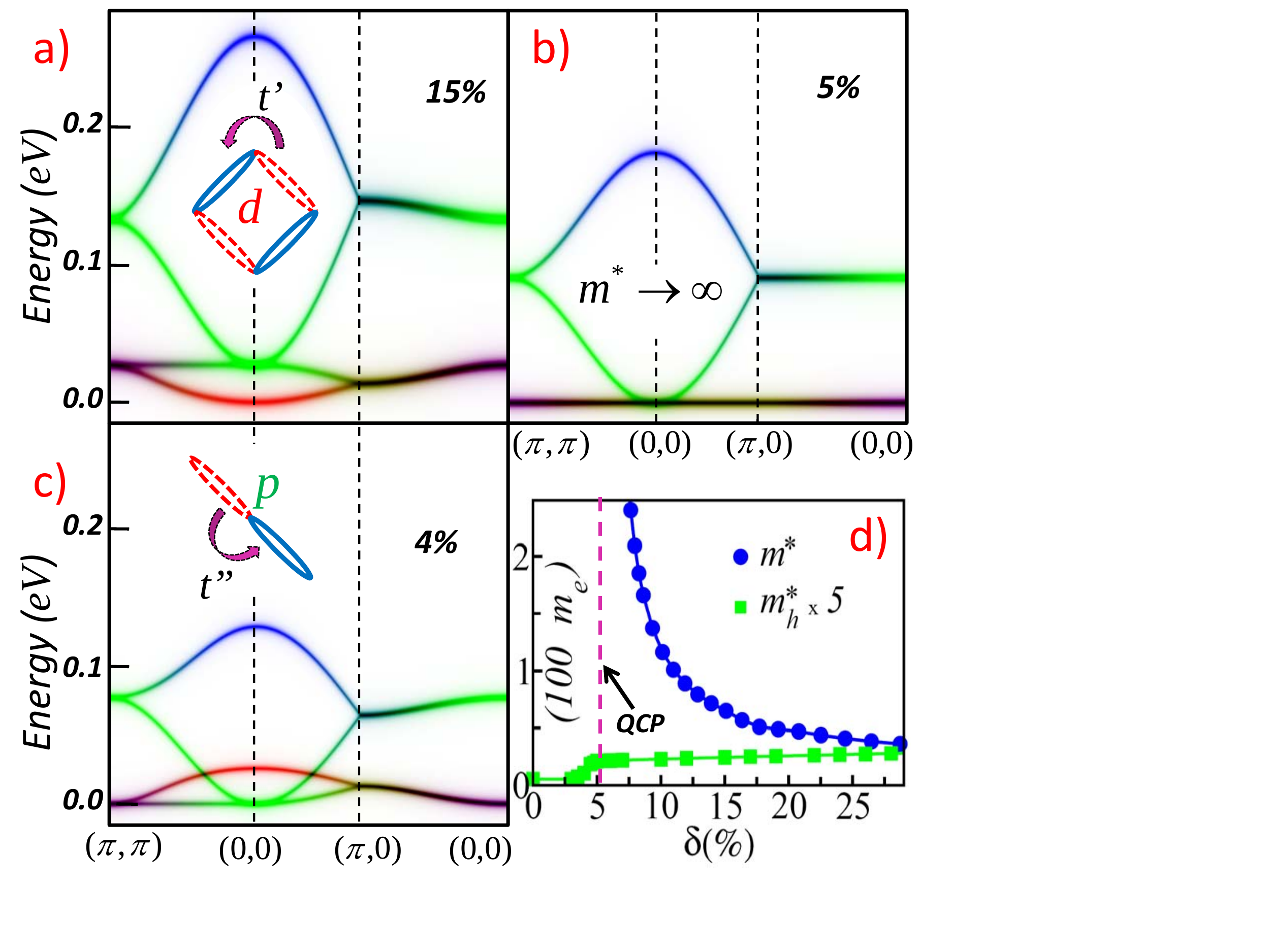}
\vskip 0.3cm
\label{F1}
\end{center}
\end{figure}

\newpage
\begin{figure}[thbp]
\begin{center}
\includegraphics[width=15.5cm,clip,angle=0]{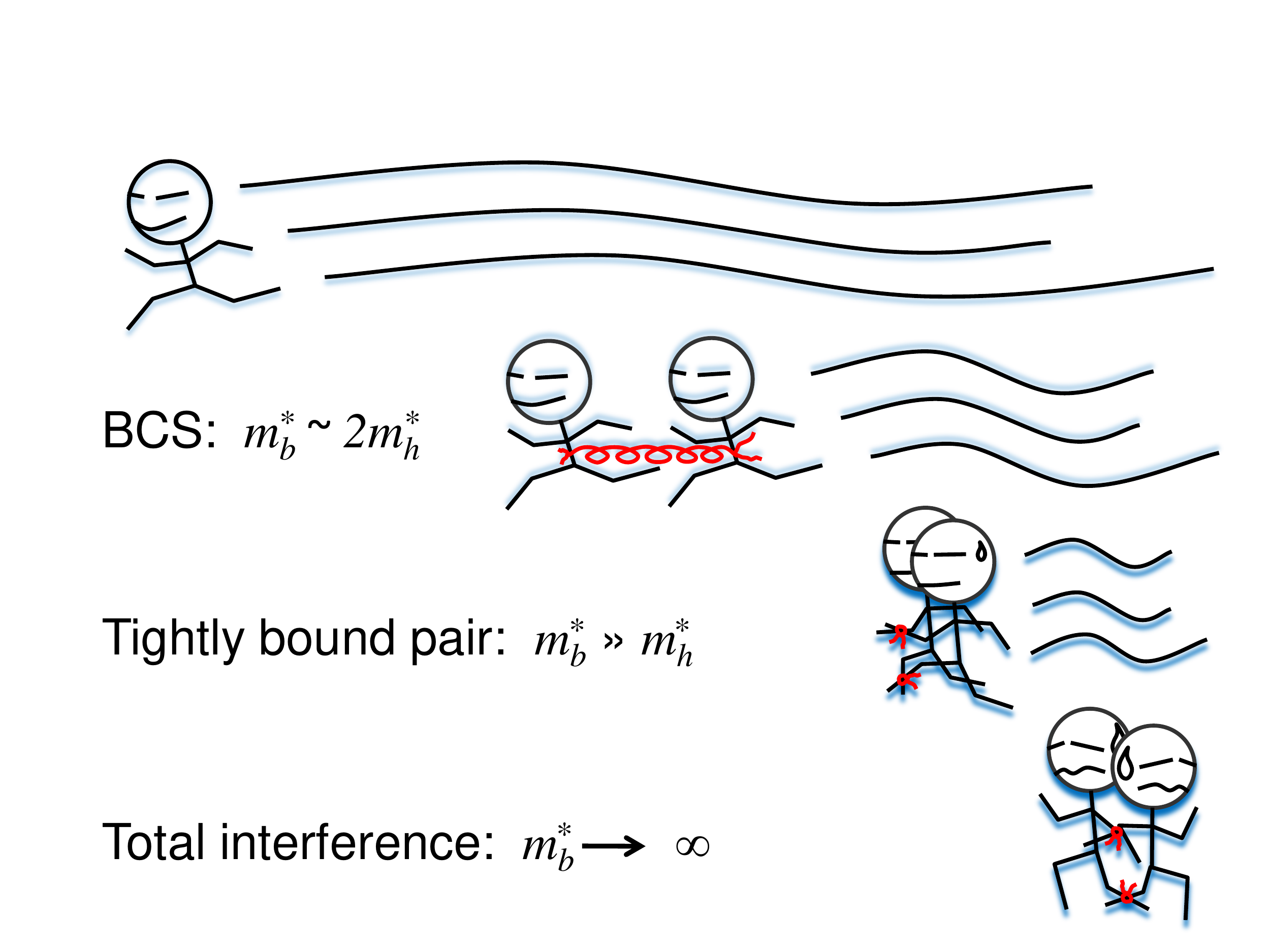}
\vskip 0.3cm
\label{F1}
\end{center}
\end{figure}

\newpage
\begin{figure}[thbp]
\begin{center}
\includegraphics[width=18.5cm,clip,angle=0]{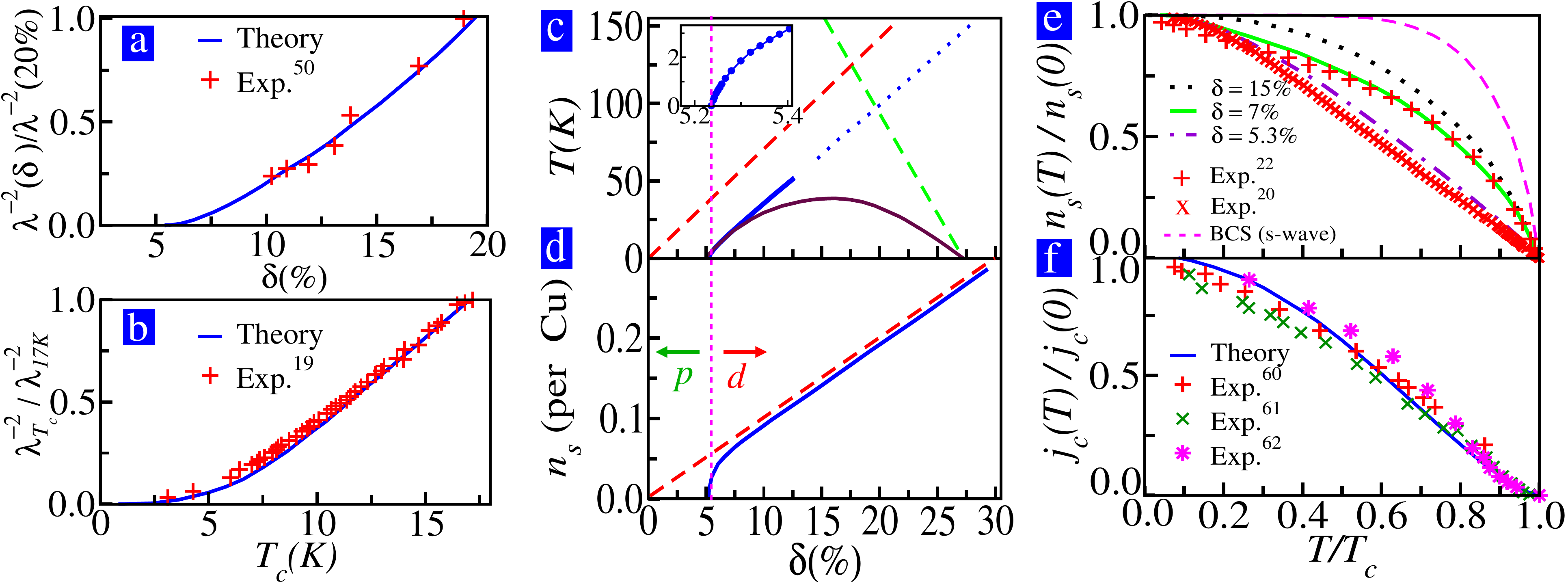}
\vskip 0.3cm
\label{F1}
\end{center}
\end{figure}


\begin{thebibliography}{}

\bibitem{EK} V.J. Emery, V.\ J.\  \&  Kivelson, S.\ A.\  Importance of phase fluctuations in superconductors with small superfluid density {\it Nature} {\bf 374}, 434--437 (1995).

\bibitem{Lee-Wen} Lee, P.\ A.\ \& Wen, X.\ G.\ Unusual Superconducting State of Underdoped Cuprates {\it Phsy.\ Rev.\ Lett.\ } {\bf 78}, 4111--4114 (1997).

\bibitem{UD1} Loeser, A.\ G.\ {\it at al.} Excitation gap in the normal state of underdoped $Bi_{2}Sr_{2}CaCu_{2}O_{8+x}$ {\it Science} {\bf 273}, 325--329 (1996).

\bibitem{UD2} Harris, J.\ M.\ {\it at al.} Anomalous superconducting state gap size versus $T_c$ behavior in underdoped $Bi_{2}Sr_{2}Ca_{1-x}Dy_{x}Cu_{2}O_{8+\delta}$ {\it Phys.\ Rev.\ B} {\bf 54}, 15665--15668 (1996).

\bibitem{UD3} Norman, M.\ R.\ {\it at al.} Destruction of the Fermi surface in underdoped highh-$T_c$ superconductors {\it Nature} {\bf 392}, 157--160 (1998).

\bibitem{UD4} Renner, Ch.\ {\it at al.} Pseudogap precursor of the superconducting gap in under-and overdoped $Bi_{2}Sr_{2}CaCu_{2}O_{8+\delta}$ {\it Phys.\ Rev.\ Lett.\ } {\bf 80}, 149--152 (1998).

\bibitem{UD5} Corson, J.\ {\it at al.} Vanishing of phase coherence in underdoped $Bi_{2}Sr_{2}CaCu_{2}O_{8+\delta}$ {\it Nature} {\bf 398}, 221--223 (1999).

\bibitem{UD6} Timusk, R.\ and Statt, B.\ The pseudogap in high-temperature superconductors: an experimental survey {\it Rep.\ Prog.\ Phys.\ } {\bf 62}, 61 --122 (1999).

\bibitem{UD7} Orenstein, J.\ \& Millis, A.\ J.\ Advances in the Physics of High-Temperature Superconductivity {\it Science} {\bf 288}, 3468--474 (2000).

\bibitem{UD8} Damascelli, A.\ , Shen, Z.\ X.\ \& Hussain, Z\. Angle-resolved photoemission studies of the cuprate superconductors. {\it Rev. Mod. Phys.} {\bf 75}, 473--541 (2003).

\bibitem{UD9} Campuzano, J.\ C.\ , Norman, M.\ R.\ \& Randeria, M.\ In the Physics of Superconductors Vol.II (eds Bennemann, K.\ H.\ \& Ketterson, J.\ B.\ )167--273 (Springer,2004).

\bibitem{UD10} Le Tacon, M.\ {\it at al.} Two energy scales and and two distinct quasiparticle dynamics in the superconducting state of underdoped cuprates {\it Nature} {\bf 2}, 537--543 (2006).

\bibitem{toni} Valla, T.\, Fedorov, A.\ V.\ , Jinho Lee, Davis, J.\ C.\ \& Gu G.\ D.\ The Ground State of the Pseudogap in Cuprate Superconductors {\it Science} {\bf 315}, 1914--1916 (2007).

\bibitem{BCS} Schrieffer, J.\ R.\ {\it Theory of Superconductivity} (Benhamin, New York, 1964).

\bibitem{Uemura} Uemura, Y.\ J.\  {\it et al.} Universal Correlations between Tc and ns / m* (Carrier Density over Effective Mass) in High-Tc Cuprate Superconductors {\it Phys.\ Rev.\ Lett.\ } {\bf 62}, 2317--2320 (1989).



\bibitem{orderpar1} Roddick, E.\ \&  Stroud D.\  Effect of Phase Fluctuations on the Low-Temperature Penetration Depth of High- Tc Superconductors {\it Phys.\ Rev.\ Lett.\ } {\bf 74}, 1430--1433 (1995).
\bibitem{orderpar2} Carlson, E.\ W.\ , Kivelson, S.\ A.\ , Emery, V.\ J.\ \&  Manousakis, E.\ Classical Phase Fluctuations in High Temperature Superconductors {\it Phys.\ Rev.\ Lett.\ } {\bf 83}, 612--615 (1999).

\bibitem{ChargedLatticeBoson} Mihlin A. and Auerbach A.\ Temperature dependence of the order parameter of cuprate superconductors {\it Phys.\ Rev.\ B\ } {\bf 80}, 134521 (2009).




\bibitem{Samuele} Samuele Sanna {\it et al.} Experimental evidence of two distinct charge carriers in underdoped cuprate superconductors {\it Phys.\ Rev.\ B} {\bf 77}, 224511 (2008).

\bibitem{Broun} Broun, D.\ M.\ {\it et al.} Superfluid Density in a Highly Underdoped $YBa_{2}Cu_{3}O_{6+y}$ Superconductor {\it Phys.\ Rev.\ Lett.\ } {\bf 99}, 237003 (2007).

\bibitem{Khasanov} Khasanov, R.\ {\it et al.} In-plane magnetic penetration depth ab in $Ca_{2-x}Na_{x}CuO_{2}Cl_{2}$: Role of the apical sites {\it Phys.\ Rev.\ B} {\bf 76}, 094505 (2007).

\bibitem{Hardy} Hardy, W.\ N.\ {\it et al.} Precision measurements of the temperature dependence of $\lambda$ in YBa$_2$Cu$_3$O$_6.95$: Strong evidence for nodes in the gap function {\it Phsy.\ Rev.\ Lett.\ } {\bf 70}, 3999--4002 (1993).


\bibitem{dia1} Yayu Wang {\it et al.} Field-Enhanced Diamagnetism in the Pseudogap State of the Cuprate $Bi_{2}Sr_{2}CaCu_{2}O_{8+\delta}$ Superconductor in an Intense Magnetic Field {\it Phys.\ Rev.\ Lett.\ } {\bf 95}, 247002 (2005).

\bibitem{dia2} Thisted, U.\ {\it et al.} Strong diamagnetic response and specific heat anomaly above $T_c$ in underdoped La$_{2-x}$Sr$_{x}$CuO$_{4}$ {\it Phys.\ Rev.\ B} {\bf 67}, 184510 (2003).

\bibitem{dia3} Iguchi, I.\ {\it et al.} Observation of diamagnetic precursor to the Meissner state above Tc in high-Tc La$_{2-x}$Sr$_x$CuO$_4$ cuprates by scanning SQUID microscopy  {\it Physica C} {\bf 367}, 9--14 (2002). 

\bibitem{pairs1} Yang, H.\ -B.\ {\it et al.} Emergence of preformed Cooper pairs from the doped Mott insulating state in $Bi{2}Sr_{2}CaCu_{2}O_{8+\delta}$ {\it Nature} {\bf 456}, 77--80 (2006).

\bibitem{pairs2} Luan, Bin-Quan  and Li, Jian-Xin  and Gong, Chang-De. Preformed pairs induced pseudogap behavior in high-$Tc$ cuprates.  {\it Phys.\ Rev.\ B} {\bf 64}, 064503 (2001).

\bibitem{RVB}  Anderson, P.\ W.\  The Resonating Valence Bond State in La2CuO4 and Superconductivity {\it Science} {\bf 235 }, 1196--1198 (1987).





\bibitem{BareMass} Cebra, B.\ and Peskin, M.\ E.\ Cooper-pair mass {\it Phys.\ Rev.\ B} {\bf 39}, 6425--6430 (1989).

\bibitem{EffectiveMass} Padilla, W.\ J.\ {\it et al} Constant effective mass across the phase digram of high-Tc cuprates {\it Phys.\ Rev.\ B} {\bf 72}, 060511(R) (2005).

\bibitem{ExpArpes1} Ino, A.\ {\it et al} Electronic structure of La$_{2-x}$Sr$_x$CuO$_4$ in the vicinity of the superconductor-insulator transition {\it Phys.\ Rev.\ B} {\bf 62}, 4137--4141 (2000).

\bibitem{ExpArpes2} Ino, A.\ {\it et al} Doping-dependent evolution of the electronic structure of La2-xSrxCuO4 in the superconducting and metallic phases{\it Phys.\ Rev.\ B} {\bf 65}, 094504 (2002).

\bibitem{ExpArpes3} Yoshida, T.\ {\it et al} Systematic doping evolution of the underlying Fermi surface of La2-xSrxCuO4 {\it Phys.\ Rev.\ B} {\bf 74}, 224510 (2006).

\bibitem{ExpArpes4} Yoshida, T.\ {\it et al} Universal versus Material-Dependent Two-Gap Behaviors in the High-Tc Cuprates: Angle-Resolved Photoemission Study of La2-xSrxCuO4. arXiv:0812.0155v1.


\bibitem{Weiguo} Wei-Guo Yin, Chang-De Gong \& Leung, P.\ W.\ Origin of the Extended Van Hove region in Cuprate Superconductors {\it Phys.\ Rev.\ Lett.\ } {\bf 81}, 2534--2537 (1998).

\bibitem{Gutzwiller} Daniel S. Rokhsar \& B. G. Kotliar Gutzwiller projection for bosons {\it Phys. Rev. B} {\bf 44}, 10328--10332 (1991).

\bibitem{biPolaron} A. Alexandrov \& Theory of Superconductivity from Weak to Strong Coupling, (Bristol and Philadelphia, 2003, 320 p).

\bibitem{AndersonGlue}  Anderson, P.\ W.\ PHYSICS: Is There Glue in Cuprate Superconductors? {\it Science } {\bf 316}, 1705--1707 (2007).

\bibitem{paper2} Yucel Yildirim and Wei Ku Unconventional superconducting gap in underdoped cuprates, arXiv:submit/0032825 (2010) {http://arxiv.org/submit/0032825}.












\bibitem{arpes2} Yoshida, T.\ {\it et al.} Metallic Behavior of Lightly Doped La$_{2-x}$Sr$_{x}$CuO$_{4}$ with a Fermi Surface Forming an Arc {\it Phys.\ Rev.\ Lett.\ } {\bf 91}, 027001 (2003).

\bibitem{arpes3} Shen, K.\ M.\ {\it et al.} Nodal Quasiparticles and Antinodal Charge Ordering in $Ca_{2-x}Na_{x}CuO_{2}Cl_{2}$ {\it Science} {\bf 307 }, 901--904 (2005).

\bibitem{arpes4} Kaminski , A.\ {\it et al.} Renormalization of Spectral Line Shape and Dispersion below $T_c$ in $Bi_{2}Sr_{2}CaCu_{2}O_{8+\delta}$  {\it Phys.\ Rev.\ Lett.\ } {\bf 86}, 1070--1073 (2001).

\bibitem{stm1}J. E. Hoffman, J.\ E.\ , McElroy, K.\ {\it et al.} Imaging Quasiparticle Interference in $Bi_{2}Sr_{2}CaCu_{2}O_{8+\delta}$ {\it Science} {\bf 297}, 1148--1151 (2002).  




\bibitem{NoPhaseSep} Beth, H.\ A.\ {\it Z. Physik} {\bf 71}, 205 (1931).



\bibitem{Elbio2} Fye, R.\ M.\, Martins, G.\ B.\ \& Dagotto, E.\  Hole-pair symmetry and excitations in the strong-coupling extended t-Jz  model: competition between d -wave and p -wave symmetry {\it Phys.\ Rev.\ B } {\bf 69}, 224507 (2004).

\bibitem{LambdaExp} Kathleen M.\ Paget {\it et al.} Magnetic penetration depth in superconducting La$_{2-x}$Sr$_{x}$CuO$_{4}$ films {\it Phys.\ Rev.\ B } {\bf 59}, 641--646 (1999).

\bibitem{OngNernstEffect&Diamagnetism} Wang, Y.\ , Li, L.\ \& Ong, N.\ P.\ Nernst effect in high-$T_c$ superconductors. {\it Phys.\ Rev.\ B} {\bf 73}, 024510 (2006).





\bibitem{MassDivMott} Casey, A.\ {\it et al.} Evidence for a Mott-Hubbard Transition in a Two-Dimensional $^3$He Fluid Monolayer {\it Phys.\ Rev.\ Lett.\ } {\bf 90}, 115301 (2003).


\bibitem{Sonier} Sonier, J.\ E.\ {\it et al.} Hope-doping dependence of the magnetic penetration depth and vortex core size in YBa$_2$Cu$_3$O$_y$: Evidence for stripe correlataions near 1/8 hole doping. {\it Phy.\ Rev.\ B} {\bf 76}, 134518 (2007).

\bibitem{Tmatrix1} Popov, V.\ N.\  Functional Integrals in Quantum Field Theory and Statistical Physics {\it Theor.\ Math.\ Phys.\ } {\bf, 11}, 565 (197). (Reidel, Dordrecht, 1983) Chap.6. p.292.

\bibitem{Tmatrix2} Hua Shi \& Allan Griffin Finite-temperature excitations in a dilute Bose-Condensed gas {\it Physics Reports} {\bf 304}, 1--87 (1998).

\bibitem{Bogoliubov} Bogoliubov, N.\ N.\ {\it Many-Body Problem} (W.\ A.\ Benjamin, New York, 1961).

\bibitem{fermionTz} Prelovsek, P.\ $c$-axis Conductivity in the Normal State of Cuprate Supercondutors  { \it Phys.\ Rev.\ Lett.\ } {\bf 81}, 3745 (1998).

\bibitem{ValueOfJ} Tokura, Y {\it et al.}  Cu-O network dependence of optical charge-transfer gaps and spin-pair excitations in single-CuO$_2$-layer compounds  { \it Phys. Rev. B} {\bf 41}, 11657 (1990).

\bibitem{AndersonLocalization} Anderson, P.\  W.\  Absence of Diffusion in Certain random Lattices  { \it Phys. Rev.} {\bf 109}, 1492 (1958).



\bibitem{quasi2D}  Bansil, A.\ , Lindroos, M.\ , Sahrakorpi,S.\ \&  Markiewicz, R.\ S.\ Influence of the third dimension of quasi-two-dimensional cuprate superconductors on angle-resolved photoemission spectra {\it Phy.\ Rev.\ B} {\bf 71}, 012503 (2005).


\bibitem{BKT1} Berezinskii, V.\ L.\ {\it Zh. Eksp. Teor. Fiz.} {\bf 61}, 1144 (1971) Engl.transl. {\it Sov.Phys.-JETP} {\bf 34}, 610 (1972).

\bibitem{BKT2} Kosterlitz, J.\ M.\ \&  Thouless, D.\ J.\ Ordering, metastability and phase transitions in two-dimensional systems {\it J. Phys. C: Solid State Phys.} {\bf 6}, 1181--1203 (1973).

\bibitem{BKT3} Kosterlitz, J.\ M. The critical properties of the two-dimensional xy model {\it J. Phys. C: Solid State Phys.} {\bf 7}, 1046--1060 (1974).


\bibitem{jc1} de Vries, J.\ W.\ C.\ , Gijs, M.\ A.\ M.\ \& Stollman, G.\ M.\ Critial current as a function of temperature in thin YBa$_2$Cu$_3$O$_{7-\delta}$ films {\it Physica C } {\bf 153-155}, 1437--1438 (1988).

\bibitem{jc2} Robertazzi, R.\ P.\ {\it et al.} n situ Ag/YBa$_2$Cu$_3$O$_7$ contacts for superconductor-normal-metal-superconductor devices {\it Phys.\ Rev.\ B} {\bf 46}, 8456--8471 (1992).

\bibitem{jc3} J.\ Mannhart, P.\ {\it et al.} Critical Currents in [001] Grains and across Their Tilt Boundaries in YBa$_2$Cu$_3$O$_7$ Films {\it Pyhs.\ Rev.\ Lett.\ } {\bf 61}, 2476--2479 (1988).

\bibitem{TranquadaStripe} Tranquada, J.\ M.\ {\it et al.} Neutron-scattering study of stripe-phase order of holes and spins in $La_{1.48}Nd_{0.4}Sr_{0.12}CuO_{4}$ {\it Phys.\ Rev.\ B} {\bf 54}, 7489--7499 (1996).


\bibitem{dDW1} Sudip Chakravarty {\it et al.} Hidden order in the cuprates {\it Phy.\ Rev.\ B} {\bf 63}, 094503 (2001).


\bibitem{dDW2} Erez Berg, Eduardo Fradkin, \& Steven A.\ Kivelson Theory of the striped superconductor {\it Phy.\ Rev.\ B} {\bf 79}, 064515 (2009).


\bibitem{STM_C2} Kohsaka, Y.\ {\it et al.} An Intrisic Bond-Centered Electronic Glass with Unidirectional Domain in Underdoped Cuprates {\it Science} {\bf 315}, 1380--1385 (2007).











\end{thebibliography}
\end{document}